\documentclass[onecolumn,unpublished,a4paper]{quantumarticle}
\usepackage{amsmath}
\usepackage{amssymb}
\usepackage{amsthm}
\usepackage{amsfonts}
\usepackage[caption=false]{subfig}
\usepackage[colorlinks]{hyperref}
\usepackage[all]{hypcap}
\usepackage{tikz}
\usepackage{relsize}
\usepackage{color,soul}
\usepackage[utf8]{inputenc}
\usepackage{capt-of}
\usepackage{mathtools}
\usepackage[numbers,sort&compress]{natbib}
\usepackage{float}
\usepackage[section]{placeins}
\usepackage{listings}
\usepackage[T1]{fontenc}  
\usetikzlibrary{decorations.pathreplacing}
\usepackage{braket}

\DeclareFixedFont{\ttb}{T1}{txtt}{bx}{n}{5}
\DeclareFixedFont{\ttm}{T1}{txtt}{m}{n}{5}
\definecolor{deepblue}{rgb}{0,0,0.5}
\definecolor{deepred}{rgb}{0.6,0,0}
\definecolor{deepgreen}{rgb}{0,0.5,0}
\newcommand\cppstyle{\lstset{
language=C++,
basicstyle=\ttm,
otherkeywords={uint8_t, __m256i, size_t, ASSERT_TRUE, EXPECT_TRUE, TEST, BENCHMARK},
keywordstyle=\ttb\color{deepblue},
emphstyle=\ttb\color{deepblue},
stringstyle=\color{deepgreen},
commentstyle=\fontfamily{txtt}\selectfont\color{gray},
showstringspaces=false,
literate={*}{{\char42}}1
         {-}{{\char45}}1
}}
\lstnewenvironment{cpp}[1][]
{\cppstyle\lstset{#1}}{}

\newcommand\pythonstyle{\lstset{
language=python,
basicstyle=\ttm,
morekeywords={assert,as,echo},
keywordstyle=\ttb\color{deepblue},
emphstyle=\ttb\color{deepblue},
stringstyle=\color{deepgreen},
commentstyle=\fontfamily{txtt}\selectfont\color{gray},
showstringspaces=false,
literate={*}{{\char42}}1
         {-}{{\char45}}1
}}
\lstnewenvironment{python}[1][]
{\pythonstyle\lstset{#1}}{}

\lstdefinestyle{stimcircuit}{
    language=python,
    basicstyle=\fontsize{4}{4}\selectfont\ttfamily,
    upquote=true,
    stepnumber=1,
    numbersep=8pt,
    showstringspaces=false,
    breaklines=true,
    frame=single,
    aboveskip=1.5em,
    belowskip=1.5em,
    commentstyle=\color{gray},
    classoffset=1,
    morekeywords={DETECTOR,OBSERVABLE_INCLUDE,rec},
    keywordstyle=\color{deepgreen},
    classoffset=2,
    morekeywords={H,R,MPP,M,MX,RX,REPEAT},
    keywordstyle=\color{deepblue},
    classoffset=3,
    morekeywords={X_ERROR,Z_ERROR,DEPOLARIZE2,DEPOLARIZE1},
    keywordstyle=\color{red},
    classoffset=4,
    morekeywords={TICK,SHIFT_COORDS,QUBIT_COORDS},
    keywordstyle=\color{gray}
}

\renewcommand\thesection{\arabic{section}}

\theoremstyle{definition}

\theoremstyle{definition}

\theoremstyle{definition}

\newcommand{\eq}[1]{\hyperref[eq:#1]{Equation~\ref*{eq:#1}}}
\renewcommand{\sec}[1]{\hyperref[sec:#1]{Section~\ref*{sec:#1}}}
\DeclareRobustCommand{\app}[1]{\hyperref[app:#1]{Appendix~\ref*{app:#1}}}
\newcommand{\fig}[1]{\hyperref[fig:#1]{Figure~\ref*{fig:#1}}}
\newcommand{\tbl}[1]{\hyperref[tbl:#1]{Table~\ref*{tbl:#1}}}
\newcommand{\theoremref}[1]{\hyperref[theorem:#1]{Theorem~\ref*{theorem:#1}}}
\newcommand{\definitionref}[1]{\hyperref[definition:#1]{Definition~\ref*{definition:#1}}}

\begin{document}
\title{A Pair Measurement Surface Code on Pentagons}

\date{\today}
\author{Craig Gidney}
\email{craig.gidney@gmail.com}
\affiliation{Google Quantum AI, Santa Barbara, California 93117, USA}

\begin{abstract}
In this paper, I present a way to compile the surface code into two-body parity measurements ("pair measurements"), where the pair measurements run along the edges of a Cairo pentagonal tiling.
The resulting circuit improves on prior work by Chao et al. by using fewer pair measurements per four-body stabilizer measurement (5 instead of 6) and fewer time steps per round of stabilizer measurement (6 instead of 10).
Using Monte Carlo sampling, I show that these improvements increase the threshold of the surface code when compiling into pair measurements from $\approx 0.2\%$ to $\approx 0.4\%$, and also that they improve the teraquop footprint at a $0.1\%$ physical gate error rate from $\approx6000$ qubits to $\approx3000$ qubits.
However, I also show that the teraquop footprint of Chao et al's construction improves more quickly than mine as physical error rate decreases, and is likely better below a physical gate error rate of $\approx 0.03\%$ (due to bidirectional hook errors in my construction).
I also compare to the planar honeycomb code, showing that although this work does noticeably reduce the gap between the surface code and the honeycomb code (when compiling into pair measurements), the honeycomb code is still more efficient (threshold $\approx 0.8\%$, teraquop footprint at $0.1\%$ of $\approx 1000$).
\end{abstract}

\maketitle

\emph{The source code that was written, the exact noisy circuits that were sampled, and the statistics that were collected as part of this paper are available at \href{https://doi.org/10.5281/zenodo.6626417}{doi.org/10.5281/zenodo.6626417}~\cite{craig_gidney_2022_6626417}.}

\section{Introduction}
\label{sec:introduction}

Last year, Hastings and Haah introduced a class of quantum error correcting codes now called "floquet codes"~\cite{hastings2021dynamically}.
Floquet codes are interesting for a lot of reasons, but from a practical perspective they're interesting because they're extremely efficient when implemented using two-body parity measurements (hereafter "pair measurements").
For example, the very first floquet code was the honeycomb code.
When using pair measurements, the honeycomb code has a threshold around $1\%$ and a teraquop footprint of around 1000 qubits at a $0.1\%$ physical gate error rate~\cite{gidney2022planarhoneycomb,paetznick2022floquetmajoranaperformance,chao2020optimization}.
(The teraquop footprint is the number of physical qubits required to create a logical qubit reliable enough to survive one trillion operations.)

From the perspective of hardware designed around pair measurements~\cite{paetznick2022floquetmajoranaperformance}, a notable problem comes up when attempting to compare floquet codes to previous work: most previous work wasn't done in terms of pair measurements.
It's far more common to see circuits compiled into unitary interactions, like CNOTs or CZs.
Hardware designed around pair measurements may not even be able to execute those circuits.
So, ideally, we'd prefer to compare to previous work that also used pair measurements.

An example of previous work using pair measurements is Chao et al.~\cite{chao2020optimization}.
They compiled the surface code into pair measurements.
The surface code is well known as one of the most promising quantum error correcting codes for building a scalable fault tolerant quantum computer, because of its high threshold and planar connectivity requirements~\cite{fowler2012surfacecodereview}.
This suggests that compiling the surface code into pair measurements should set a solid baseline that other codes can compare against.
However, ignoring the problems with comparing across gate sets for the moment, Chao et al. found that the surface code performed much worse when using pair measurements.
Their pair measurement surface code's threshold was roughly $0.2\%$ (as opposed to roughly $0.8\%$ when using unitary interactions).
For comparison, the honeycomb code has a threshold around $1\%$ when using pair measurements~\cite{gidney2022planarhoneycomb,paetznick2022floquetmajoranaperformance}.

This paper is my attempt to improve on Chao et al's result.
There are three reasons I thought it would be possible to do better.
First, Chao et al's work conceptualizes the problem as a quantum circuit problem.
In my experience, the ZX calculus~\cite{coecke2011introducezx} is a better tool for reasoning about and optimizing measurement based constructions~\cite{de2017zxlattice}.
I suspected that if I modelled the problem as a ZX graph, I could find a better solution.
Second, Chao et al. used a union find decoder with unweighted edges~\cite{delfosse2021almost}.
By contrast, estimates of the cost of the honeycomb code have been based on correlated minimum weight perfect matching~\cite{gidney2021honeycombmemory,gidney2022planarhoneycomb}, which is slower but more accurate.
Using the same decoder for the surface code would make the comparison fairer.
Third, Chao et al. only estimated the threshold.
I wanted to check if the loss in threshold was actually matched by an increase in the teraquop footprint.

The paper is divided as follows.
In \sec{construction}, I present the more compact construction that I found by using the ZX calculus.
In \sec{results}, I present the simulation data obtained using a correlated minimum weight perfect matching decoder.
In \sec{conclusion}, I discuss conclusions, including the ironic fact that although  my construction has a better threshold, its teraquop footprint eventually becomes larger than Chao et al.'s as physical error rate decreases.
The paper also includes \app{uncertainty}, which specifies how I computed line fits, \app{noise_model}, which specifies the noise model I used, and \app{other_plots}, which includes additional data plots.

\begin{figure}
    \centering
    \resizebox{0.99\textwidth}{!}{
        \includegraphics{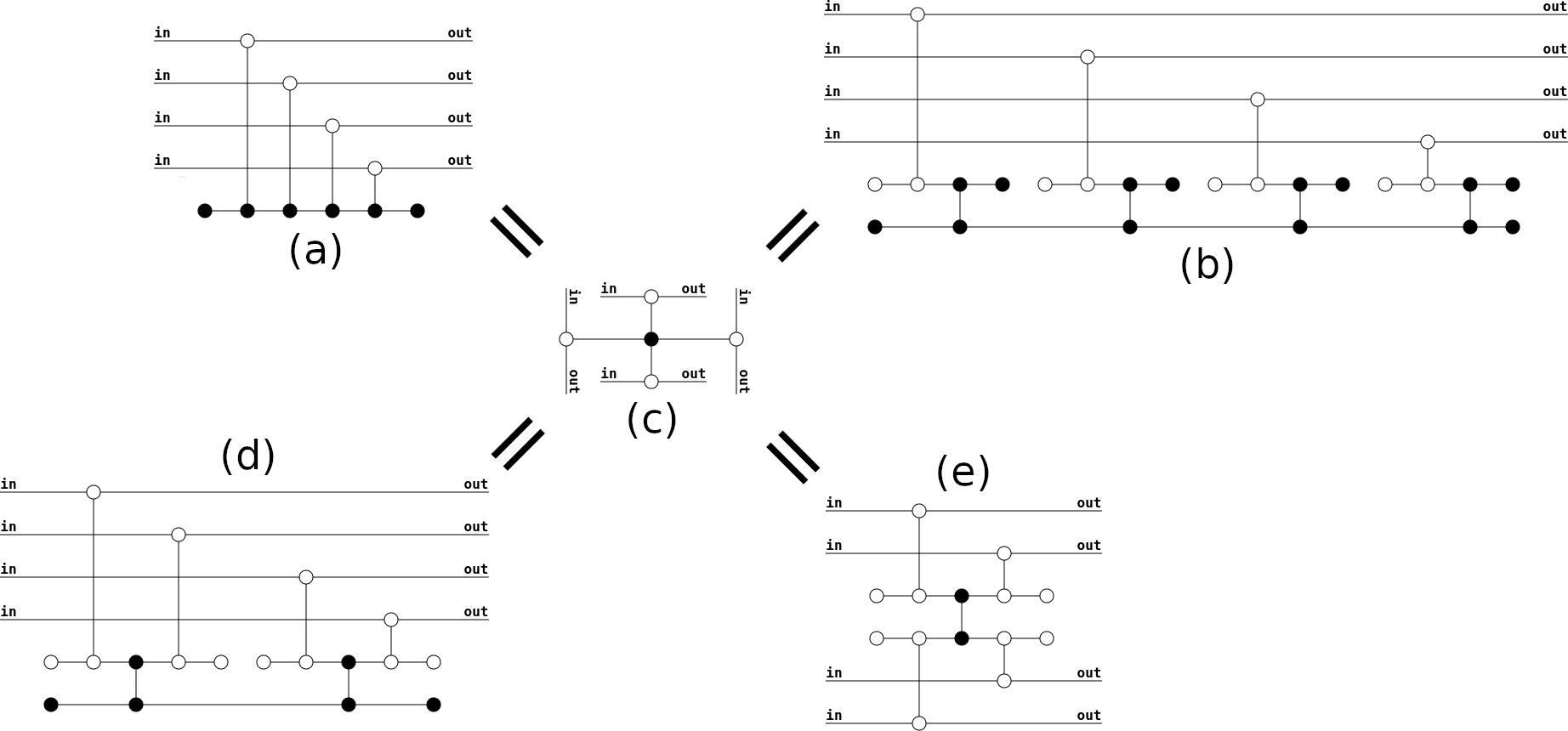}
    }
    \caption{
        Different ZX graphs implementing a four-body parity measurement.
        (a) The ZX graph for the well known decomposition into four CNOT operations.
        (b) The ZX graph produced by replacing each CNOT with a lattice surgery CNOT~\cite{horsman2012latticesurgery}.
        (c) The smallest possible ZX graph implementing a four-body parity measurement.
        (d) The construction found by Chao et al. using brute force search~\cite{chao2020optimization}.
        (e) The construction presented by this paper.
        The graphs are all equal because, when spider fusion is repeatedly applied, they reduce to (c).
        White circles are X type nodes.
        Black circles are Z type nodes.
        Assuming time moves from left to right, leaf nodes correspond to single qubit initializations and demolition measurements while vertical edges correspond to CNOT operations (if the linked nodes have opposite type) or pair measurements (if the linked nodes have the same type).
    }
    \label{fig:zx_identities}
\end{figure}

\section{Construction}
\label{sec:construction}

This paper is not intended to be an introduction to the ZX calculus.
I'll cover barely enough detail to convey that an efficient graph that implements a four-body X basis parity measurement is going to look like a Z type core surrounded by X type limbs leading to the data qubits.
I recommend any of \cite{backens2016simplifiedzx,de2017zxlattice,coecke2017picturing} as starting points for learning the ZX calculus.
Note that I will be using a non-standard node coloring, where X nodes are white (instead of red) and Z nodes are black (instead of green).
I use these colors because it makes the ZX graph of a CNOT gate look like the quantum circuit diagram of a CNOT gate.

In the ZX calculus, an X basis parity measurement can be added to a graph by placing an X type node on each involved qubit and linking all of the nodes to a central Z type node (see graph (c) of \fig{zx_identities}).
To create other implementations of this parity measurement, rewrite rules can be applied to the graph~\cite{backens2016simplifiedzx}.
For example, the "spider fusion" rule (called "S1" in \cite{backens2016simplifiedzx}) allows any edge to be contracted if it's between nodes of the same type.
As another example, the "redundant node" rule (called "S3" in \cite{backens2016simplifiedzx}) allows nodes to be added to the middle of edges.

Consider the following circuit decomposition of an X basis parity measurement.
First, init an ancilla in the X basis producing a $|+\rangle$.
Then, for each involved data qubit, apply a CNOT controlled by the ancilla and targeting the data qubit.
Finally, demolition measure the ancilla in the X basis.
In the ZX calculus, a CNOT is represented by a Z type node on the control linked to an X type node on the target.
X basis init and demolition measurement are both represented by Z type leaf nodes.
(It's somewhat confusing that the Z type leaf prepares X type states, but this choice is a natural consequence of how the nodes are defined and avoids requiring special cases for leaves in the rewrite rules.)
So, after converting into a ZX graph, the circuit construction would look like a Z type leaf (the init) leading into a series of Z type nodes linked to X type nodes on the data qubits (the CNOTs), leading into a Z type leaf (the measurement).
This graph is graph (a) in \fig{zx_identities}.
By repeatedly applying the spider fusion rule, you can contract all of the Z type nodes into a single node, producing graph (c) in \fig{zx_identities}, proving the circuit decomposition is correct.

Now consider the second circuit shown in Figure 6 of Chao et al's paper~\cite{chao2020optimization}.
This circuit involves single-qubit non-demolition measurements, X pair measurements, Z pair measurements, and Pauli feedback.
The Pauli feedback is irrelevant in the ZX calculus, and can be added in again later, so the first step to converting the circuit into a ZX graph is to delete all of the feedback.
Each X basis pair measurement becomes a pair of linked X type nodes (normally an X parity measurement would link the qubits to a central Z type node but, in the specific case of a pair measurement, that node can be omitted because of the "redundant node" rule).
Similarly, Z basis pair measurements become a pair of linked Z type nodes.
Each single qubit measurement ends up being a leaf node, via rule B1 from \cite{backens2016simplifiedzx}, with some of the leaf nodes corresponding to demolition measurements and some corresponding to inits.
The final converted graph is graph (d) in \fig{zx_identities}.
You can verify that it's correct by repeatedly applying the spider fusion rule.

\begin{figure}
    \centering
    \resizebox{0.99\textwidth}{!}{
        \includegraphics{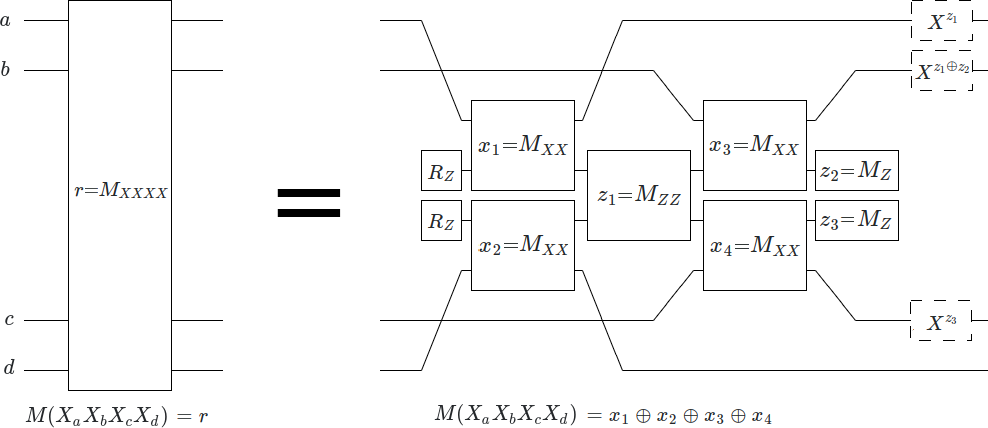}
    }
    \caption{
        Decomposition of a four-body parity measurement into five pair measurements.
        This circuit was derived from graph (e) in \fig{zx_identities}.
        Dashed operations are classical Pauli feedback, which can be removed by folding its effects into the definitions of detectors and observables used by the surrounding surface code (see \fig{detector}).
        \href{https://algassert.com/quirk\#circuit=\%7B\%22cols\%22\%3A\%5B\%5B\%22~evn6\%22\%2C\%22~evn6\%22\%2C1\%2C1\%2C\%22~evn6\%22\%2C\%22~evn6\%22\%5D\%2C\%5B1\%2C1\%2C\%22~jqn9\%22\%2C\%22~jqn9\%22\%2C1\%2C1\%2C\%22~7jvd\%22\%2C\%22~7jvd\%22\%5D\%2C\%5B\%22X\%22\%2C1\%2C\%22X\%22\%2C1\%2C1\%2C1\%2C\%22\%E2\%8A\%96\%22\%5D\%2C\%5B1\%2C1\%2C1\%2C\%22X\%22\%2C1\%2C\%22X\%22\%2C1\%2C\%22\%E2\%8A\%96\%22\%5D\%2C\%5B1\%2C1\%2C1\%2C1\%2C1\%2C1\%2C\%22Measure\%22\%2C\%22Measure\%22\%5D\%2C\%5B1\%2C1\%2C1\%2C1\%2C1\%2C1\%2C1\%2C1\%2C\%22~7jvd\%22\%5D\%2C\%5B1\%2C1\%2C\%22Z\%22\%2C\%22Z\%22\%2C1\%2C1\%2C1\%2C1\%2C\%22\%E2\%8A\%96\%22\%5D\%2C\%5B1\%2C1\%2C1\%2C1\%2C1\%2C1\%2C1\%2C1\%2C\%22Measure\%22\%5D\%2C\%5B1\%2C1\%2C1\%2C1\%2C1\%2C1\%2C1\%2C1\%2C1\%2C\%22~7jvd\%22\%2C\%22~7jvd\%22\%5D\%2C\%5B1\%2C\%22X\%22\%2C\%22X\%22\%2C1\%2C1\%2C1\%2C1\%2C1\%2C1\%2C\%22\%E2\%8A\%96\%22\%5D\%2C\%5B1\%2C1\%2C1\%2C\%22X\%22\%2C\%22X\%22\%2C1\%2C1\%2C1\%2C1\%2C1\%2C\%22\%E2\%8A\%96\%22\%5D\%2C\%5B1\%2C1\%2C1\%2C1\%2C1\%2C1\%2C1\%2C1\%2C1\%2C\%22Measure\%22\%2C\%22Measure\%22\%5D\%2C\%5B1\%2C1\%2C\%22Measure\%22\%2C\%22Measure\%22\%5D\%2C\%5B1\%2C\%22X\%22\%2C\%22\%E2\%80\%A2\%22\%5D\%2C\%5B1\%2C1\%2C1\%2C\%22\%E2\%80\%A2\%22\%2C\%22X\%22\%5D\%2C\%5B\%22X\%22\%2C\%22X\%22\%2C1\%2C1\%2C1\%2C1\%2C1\%2C1\%2C\%22\%E2\%80\%A2\%22\%5D\%2C\%5B1\%2C1\%2C1\%2C1\%2C1\%2C1\%2C\%22\%E2\%80\%A2\%22\%2C1\%2C1\%2C1\%2C\%22X\%22\%5D\%2C\%5B1\%2C1\%2C1\%2C1\%2C1\%2C1\%2C1\%2C\%22\%E2\%80\%A2\%22\%2C1\%2C1\%2C\%22X\%22\%5D\%2C\%5B1\%2C1\%2C1\%2C1\%2C1\%2C1\%2C1\%2C1\%2C1\%2C\%22\%E2\%80\%A2\%22\%2C\%22X\%22\%5D\%2C\%5B\%22~jcpu\%22\%2C\%22~jcpu\%22\%2C1\%2C1\%2C\%22~jcpu\%22\%2C\%22~jcpu\%22\%2C1\%2C1\%2C1\%2C1\%2C\%22~s93v\%22\%5D\%2C\%5B\%22~jjav\%22\%2C\%22~jjav\%22\%2C\%22~jjav\%22\%2C\%22~jjav\%22\%2C\%22~jjav\%22\%2C\%22~jjav\%22\%2C\%22~jjav\%22\%2C\%22~jjav\%22\%2C\%22~jjav\%22\%2C\%22~jjav\%22\%2C\%22~jjav\%22\%2C\%22~jjav\%22\%5D\%2C\%5B1\%2C1\%2C\%22\%3C\%3C3\%22\%5D\%2C\%5B1\%2C1\%2C1\%2C\%22\%3C\%3C3\%22\%5D\%2C\%5B\%22Amps4\%22\%2C1\%2C1\%2C1\%2C1\%2C1\%2C1\%2C1\%2C1\%2C1\%2C\%22\%E2\%97\%A6\%22\%5D\%2C\%5B\%5D\%2C\%5B\%5D\%2C\%5B\%5D\%2C\%5B\%22Amps4\%22\%2C1\%2C1\%2C1\%2C1\%2C1\%2C1\%2C1\%2C1\%2C1\%2C\%22\%E2\%80\%A2\%22\%5D\%2C\%5B\%5D\%2C\%5B\%5D\%2C\%5B\%5D\%2C\%5B\%22X\%22\%2C\%22X\%22\%2C\%22X\%22\%2C\%22X\%22\%2C1\%2C1\%2C1\%2C1\%2C1\%2C1\%2C1\%2C\%22\%E2\%8A\%96\%22\%5D\%2C\%5B1\%2C1\%2C1\%2C1\%2C1\%2C1\%2C1\%2C1\%2C1\%2C1\%2C\%22\%E2\%80\%A2\%22\%2C\%22X\%22\%5D\%2C\%5B1\%2C1\%2C1\%2C1\%2C1\%2C1\%2C1\%2C1\%2C1\%2C1\%2C1\%2C\%22Chance\%22\%5D\%5D\%2C\%22gates\%22\%3A\%5B\%7B\%22id\%22\%3A\%22~evn6\%22\%2C\%22name\%22\%3A\%22in\%22\%2C\%22matrix\%22\%3A\%22\%7B\%7B1\%2C0\%7D\%2C\%7B0\%2C1\%7D\%7D\%22\%7D\%2C\%7B\%22id\%22\%3A\%22~jjav\%22\%2C\%22name\%22\%3A\%22check\%22\%2C\%22matrix\%22\%3A\%22\%7B\%7B1\%2C0\%7D\%2C\%7B0\%2C1\%7D\%7D\%22\%7D\%2C\%7B\%22id\%22\%3A\%22~jcpu\%22\%2C\%22name\%22\%3A\%22out\%22\%2C\%22matrix\%22\%3A\%22\%7B\%7B1\%2C0\%7D\%2C\%7B0\%2C1\%7D\%7D\%22\%7D\%2C\%7B\%22id\%22\%3A\%22~s93v\%22\%2C\%22name\%22\%3A\%22result\%22\%2C\%22matrix\%22\%3A\%22\%7B\%7B1\%2C0\%7D\%2C\%7B0\%2C1\%7D\%7D\%22\%7D\%2C\%7B\%22id\%22\%3A\%22~7jvd\%22\%2C\%22name\%22\%3A\%22store\%22\%2C\%22matrix\%22\%3A\%22\%7B\%7B1\%2C0\%7D\%2C\%7B0\%2C1\%7D\%7D\%22\%7D\%2C\%7B\%22id\%22\%3A\%22~jqn9\%22\%2C\%22name\%22\%3A\%22\%7C0\%3E\%22\%2C\%22matrix\%22\%3A\%22\%7B\%7B1\%2C0\%7D\%2C\%7B0\%2C1\%7D\%7D\%22\%7D\%5D\%7D}{(Click here to open an equivalent circuit in Quirk.)}
    }
    \label{fig:xxxx_circuit}
\end{figure}

So far, all of these constructions look like a Z type core with X type limbs leading to each data qubit.
This is not technically required to be true, but because we're trying to find an efficient construction it's unlikely that there's space to do anything else.
So let's explicitly assume that the optimal construction will have this form.
Notice that this implies the graph must have four core-to-limb transitions, where there is an edge between a Z type node (from the core) and an X type node (from a limb).
Since pair measurements don't link nodes of opposite type, and the only interactions we're using are pair measurements, the core-to-limb transitions have to occur along timelike edges (edges that correspond to the worldline of a qubit).
Supposing the core was a single Z type pair measurement, there would be exactly four timelike edges coming out of this pair measurement (two inputs and two outputs).
That's exactly enough to get the four transitions that must appear in the graph.
Surrounding this central pair measurement with four X type pair measurements, one leading to each data qubit, is then exactly enough connections to have each limb reach a data qubit and connect the whole system together.
This construction is graph (e) of \fig{zx_identities}.
By spider fusion, it implements the desired four-body parity measurement.
The corresponding circuit (with feedback restored) is shown in \fig{xxxx_circuit}.

A different way to prove that this construction is correct is to show that it has all of the stabilizer generators of a four-body $X$ basis parity measurement.
These generators are $X_1 \rightarrow X_1$,
$X_2 \rightarrow X_2$,
$X_3 \rightarrow X_3$,
$X_4 \rightarrow X_4$,
$Z_1Z_2 \rightarrow Z_1Z_2$,
$Z_2Z_3 \rightarrow Z_2Z_3$,
$Z_3Z_4 \rightarrow Z_3Z_4$,
and $X_1 X_2 X_3 X_4 \rightarrow 1$.
\fig{zx_stabilizer_examples} gives example visual proofs for three out of these eight rules.
\fig{zx_detector_example} then shows how these rules can be chained together to form detectors.

\begin{figure}
    \centering
    \resizebox{0.5\textwidth}{!}{
        \includegraphics{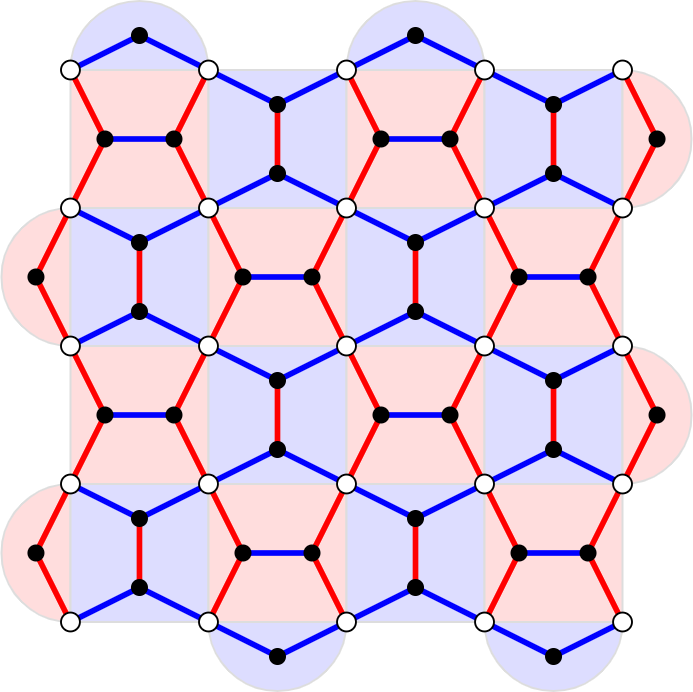}
    }
    \caption{
        Pair measurement layout for a 5x5 surface code.
        White circles are data qubits.
        Black circles are measurement qubits.
        Red shapes are the X stabilizers of the surface code.
        Blue shapes are the Z stabilizers of the surface code.
        Red edges are XX pair measurements performed by the circuit.
        Blue edges are ZZ pair measurements performed by the circuit.
        The red and blue edges form a planar graph with pentagonal faces.
    }
    \label{fig:tiling}
\end{figure}

The four body parity measurement construction that I've described uses five pair measurements.
This is fewer than the six pair measurement construction reported by Chao et al.
This is surprising, because Chao et al. ran a brute force search of the solution space.
I contacted Chao et al. to ask why their search missed the five pair measurement solution.
I had assumed it was because their search had skipped the circuit due to the bidirectional hook errors, or because it didn't follow their desired connectivity, but they explained to me that what they brute forced was how to implement a double-target CNOT gate using pair measurements.
They decomposed the four-body measurement into two double-target CNOTs, and then decomposed each double-target CNOT into the three pair measurement decomposition they had found for it, resulting in a six pair measurement construction.

A hook error is a single physical error that affects multiple data qubits due to the details of how abstract operations required by an error correcting code are decomposed into an explicit circuit.
Look closely at \fig{xxxx_circuit}.
A hook error equivalent to the data errors $Z_a Z_b$ occurs if the result of the central measurement (labelled $z_1$) is flipped.
A hook error equivalent to the data errors $Z_b Z_c$ occurs if the two qubit depolarizing noise during the central measurement applies $X \otimes X$ to the two measurement qubits.
These two different hook errors are equivalent to pairs of data errors.
These pairs overlap on one data qubit.
Therefore, when laid out in 2d, these hook errors will move in different directions.
This is a problem because error chains will be able to use these hook errors to grow at double speed both vertically and horizontally, regardless of the orientation and ordering used when interleaving parity measurements.
This cuts the code distance of the construction in half.

I'd normally consider halving the code distance to be a showstopping problem.
In fact, if I'd realized early on that my construction had bidirectional hook errors, I'd have dropped the whole project and moved on to something else.
But I actually only caught the problem later, when trying to understand my initial results, and the initial results were promising enough that I decided to continue despite the problem.

When mapping the circuit shown in \fig{xxxx_circuit} into 2d, it's natural to place the respective measurement qubits close to the two data qubits they each interact with.
The connections required by the circuit then form a ``puckered H" shape.
Although there are hook errors in both directions, it's still the case that one hook error is more likely than the other.
So, it's still beneficial to orient the puckered H layouts one way for the Z basis measurements and the other way for the X basis measurements (because the X and Z boundaries are in different directions).
\fig{tiling} shows the result of applying this to each of the stabilizers of the surface code.
Pleasingly, the resulting set of connections form a pentagonal tiling of the plane known as the Cairo pentagonal tiling~\cite{wiki:Cairo_pentagonal_tiling}.

I tried a few possible ways of interleaving the stabilizer measurements.
The best interleaving that I found was to do all of the X basis stabilizer measurements then all of the Z basis stabilizer measurements.
If these two steps are partially pipelined, then only six additional layers of circuit are needed per round of stabilizer measurements.
The exact circuit interleaving, and the resulting structure of a detector, are shown in \fig{detector}.

\begin{figure}[h]
    \centering
    \resizebox{0.7\textwidth}{!}{
        \includegraphics{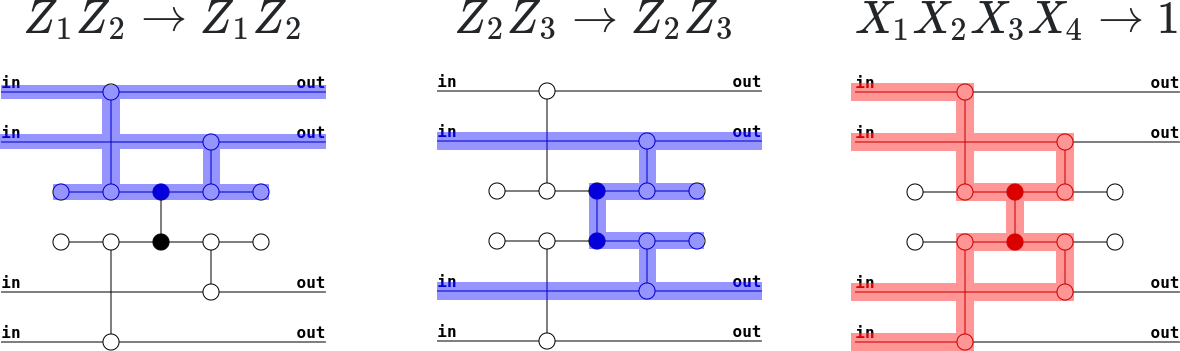}
    }
    \caption{
        Examples of stabilizer generators satisfied by a ZX graph implementing a four-body parity measurement.
        Red highlight represents X sensitivity.
        Blue highlight represents Z sensitivity.
        For highlights to be valid they must satisfy two rules.
        First, sensitivity \emph{crosses} nodes of the same type.
        All white (black) nodes must have red (blue) highlight on an even number of adjacent edges.
        Second, sensitivity \emph{broadcasts} over nodes of the opposite type.
        All black (white) nodes must have red (blue) highlight on all adjacent edges or on no adjacent edges.
    }
    \label{fig:zx_stabilizer_examples}
\end{figure}

\begin{figure}[H]
    \centering
    \resizebox{0.7\textwidth}{!}{
        \includegraphics{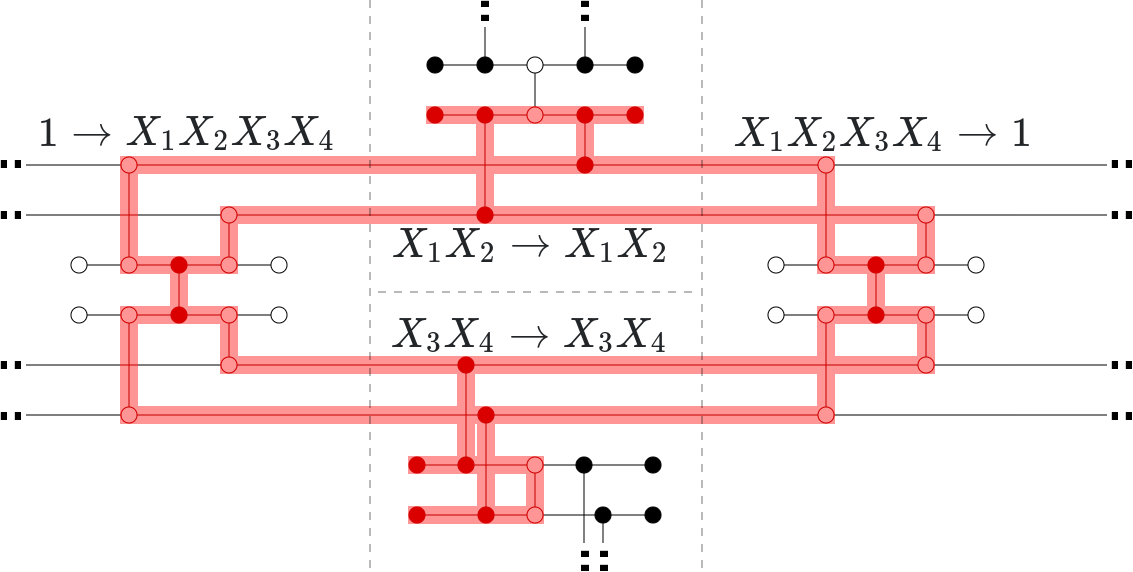}
    }
    \caption{
        Example of a detector (an internal tautology) in a ZX graph built from four-body parity measurements.
        This graph is a simplified version of what occurs in a surface code (in particular, the detector crosses two parity measurements in the opposite basis instead of four).
        When converting into a circuit, the measurements that define the detector will correspond to the highlighted pair measurements in the same basis (vertical edges between white nodes) combined with the highlighted single qubit measurements in the opposite basis (black leaf nodes pointing rightward).
    }
    \label{fig:zx_detector_example}
\end{figure}

\begin{figure}
    \centering
    \resizebox{0.99\textwidth}{!}{
        \includegraphics{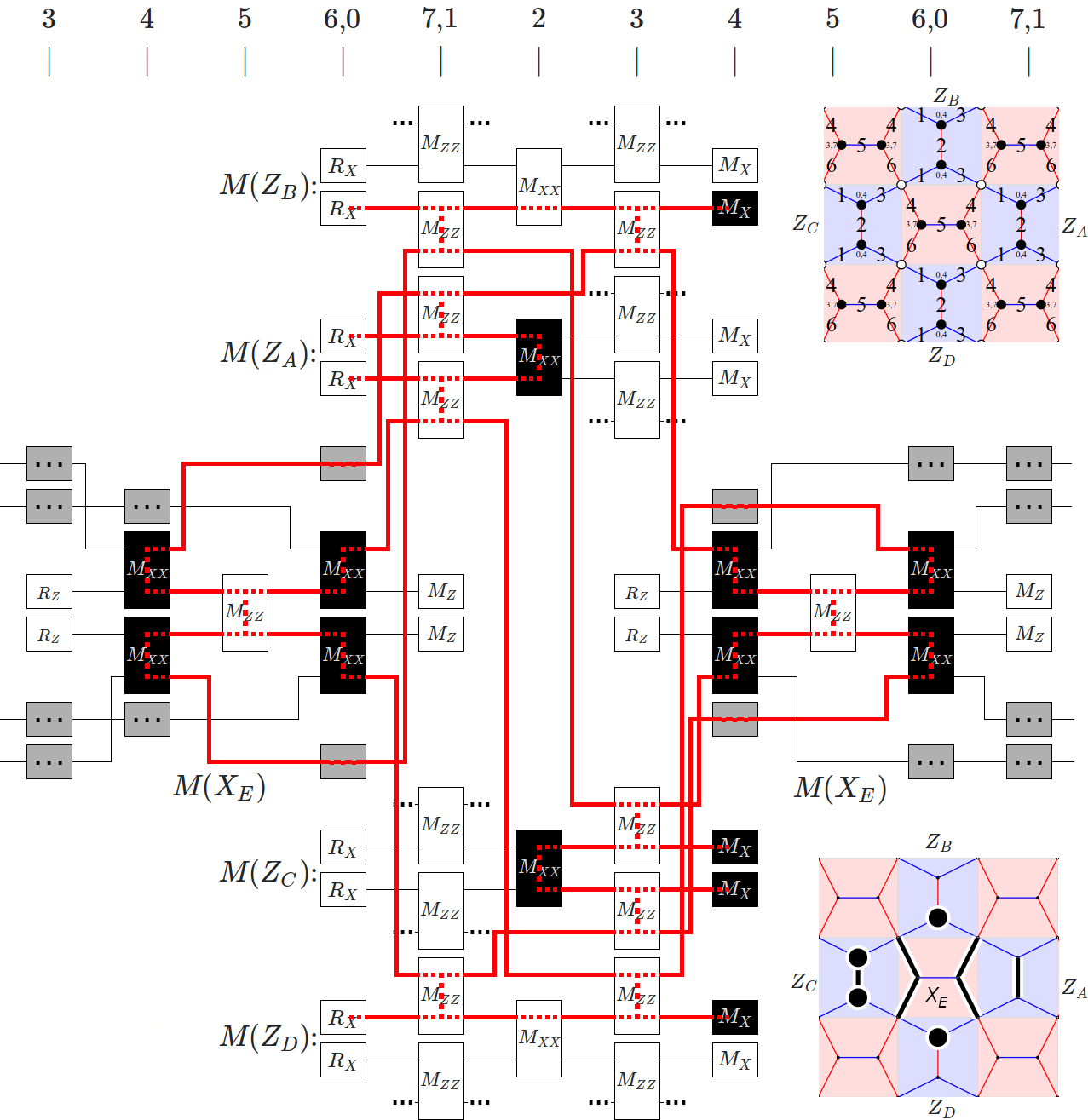}
    }
    \caption{
        Order of operations used by my construction, and the resulting structure of a detector in the bulk.
        In the circuit diagram, the parity of the 14 measurements colored black is always the same under noiseless execution, forming a detector.
        Red highlights show where a Z or Y error would cause the detector to produce a detection event.
        Gray boxes with dots always correspond to an $M_{XX}$ measurement with a qubit not shown in the circuit diagram.
        The top right panel shows the 2d layout of the circuit with layer order annotations.
        The top axis indicates the layers of the circuit diagram, corresponding to the layer annotations in the top right panel.
        The overlapping layers ("6,0" and "7,1") are due to partial pipelining of the rounds of stabilizer measurements.
        The bottom right panel shows where the measurements included in the detector are, in space, with bold lines for included pair measurements and bold circles for included single qubit measurements.
    }
    \label{fig:detector}
\end{figure}

\section{Simulation}
\label{sec:results}

To quantify how well my construction works I used Stim, Sinter, and an internal correlated minimum weight perfect matching decoder written by Austin Fowler.
Stim is a tool for fast simulation and analysis of stabilizer circuits~\cite{gidney2021stim}.
Sinter is a tool for bulk sampling and decoding of Stim circuits using python multiprocessing~\cite{sinter-source}.

I wrote python code to generate Stim circuits representing my construction, for various error rates and patch sizes.
I also implemented python code to generate Stim circuits reproducing the best construction described by Chao et al.~\cite{chao2020optimization}.
I also generated honeycomb code circuits for comparison, by using the python code attached to \cite{gidney2022planarhoneycomb}.
I added noise to these circuits using a noise model described in detail in \app{noise_model}.
I then sampled the logical error rate of each noisy circuit, targeting a hundred million shots or ten thousand errors (whichever came first).
Using a 96 core machine, the sampling process took approximately four days.

\begin{figure}[h]
    \centering
    \resizebox{0.99\textwidth}{!}{
        \includegraphics{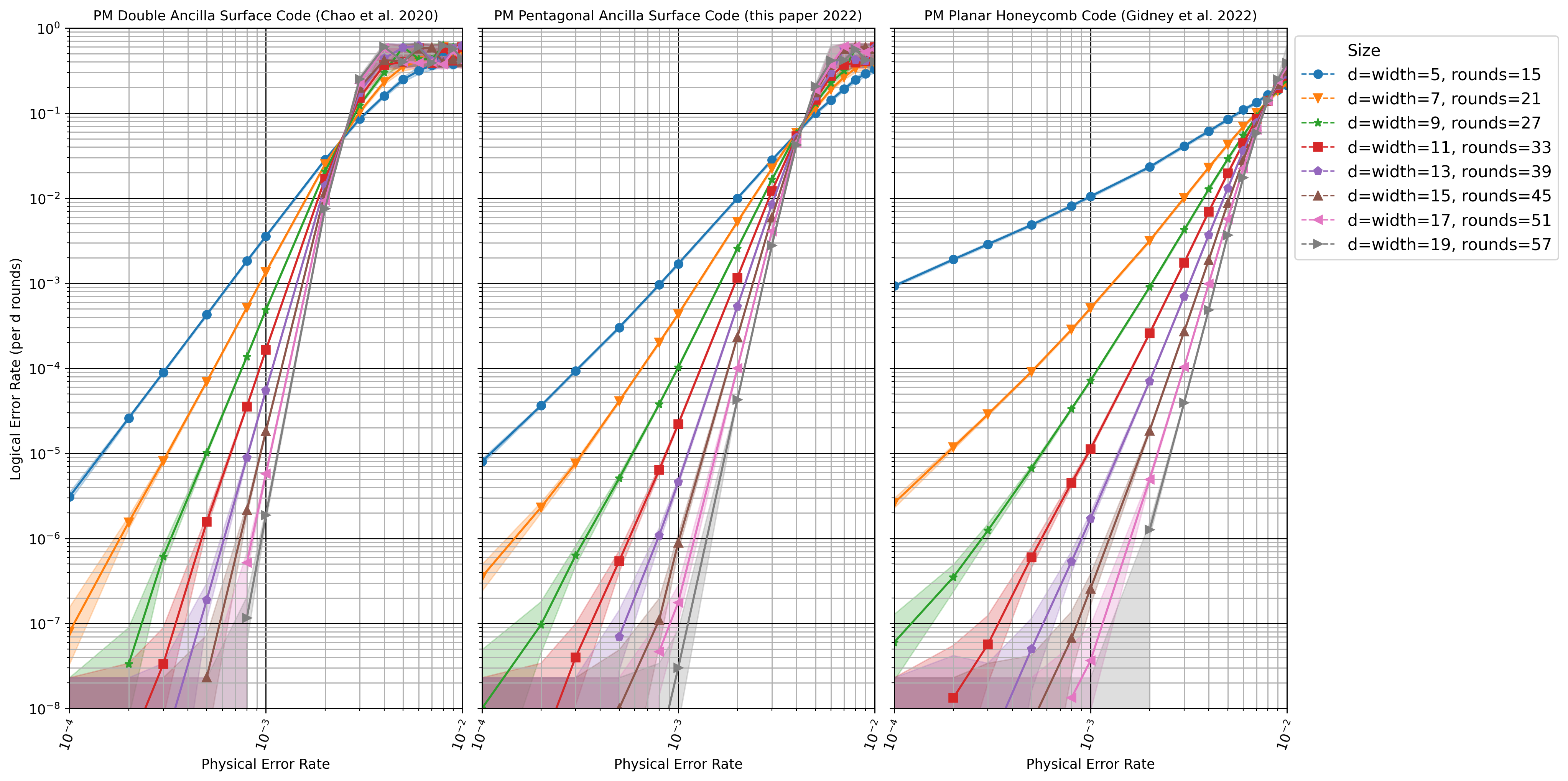}
    }
    \caption{
        Physical error rate vs logical error rate for various patch widths.
        Based on X basis memory experiments; see \app{other_plots} for Z basis.
        Note that the honeycomb code uses a qubit patch with a 2:3 aspect ratio (costs an extra factor of 1.5 in qubit count relative to the surface code constructions) but has no measurement ancillae (saves a factor of 1.5 in qubit count relative to the surface code constructions), so ultimately all the constructions have similar qubit counts at a given width.
        Decoding was done using correlated minimum weight perfect matching.
        Shading shows the range of hypothesis probabilities with a likelihood within a factor of 1000 of the max likelihood hypothesis, given the sampled data.
    }
    \label{fig:error_rate_plot}
\end{figure}

\fig{error_rate_plot} shows the sampled logical error rate from each construction for various code distances and physical error rates.
It shows that the threshold of my construction is roughly double the threshold of Chao et al.'s construction.
However, if you look closely at the slopes of the curves, you can see the effects of the bidirectional hook errors: my construction is improving slower than Chao et al's as the physical error rate decreases.

\begin{figure}[h]
    \centering
    \resizebox{0.99\textwidth}{!}{
        \includegraphics{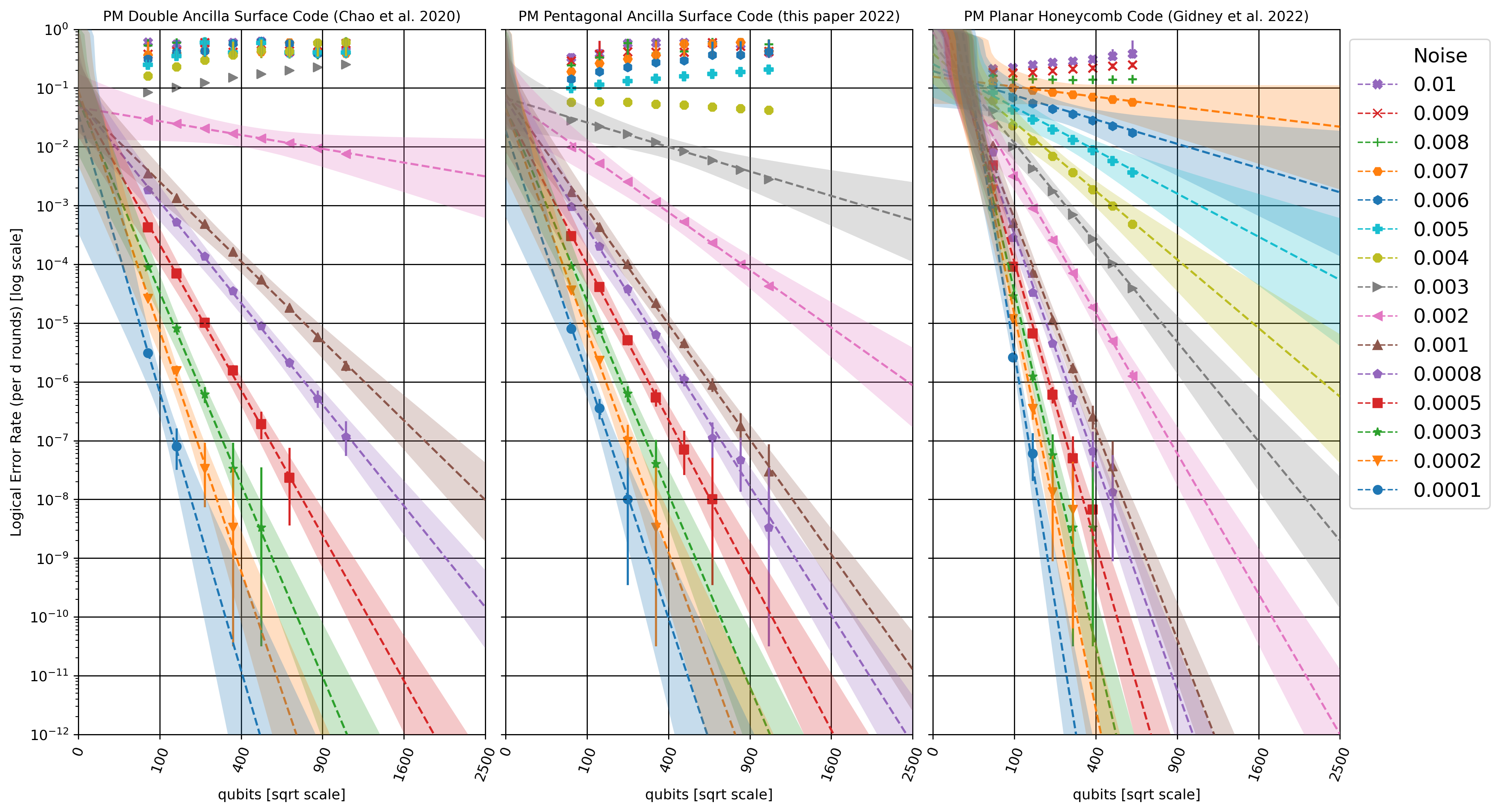}
    }
    \caption{
        Linear extrapolation of log logical error rate versus square root qubit count, for various physical error rates.
        Based on X basis memory experiments; see \app{other_plots} for Z basis.
        The vertical bar attached to each point shows the range of hypothesis probabilities with a likelihood within a factor of 1000 of the max likelihood hypothesis, given the sampled data.
        See \app{uncertainty} for a discussion of how the line fits were computed.
    }
    \label{fig:extrapolation_plot}
\end{figure}

\fig{extrapolation_plot} shows essentially the same data as \fig{error_rate_plot}, but accompanied by line fits that project the number of qubits required to achieve a given logical error rate.
The line fits were computed using a Bayesian method, with a truncation step to represent systemic uncertainty (see \app{uncertainty} for details).

\begin{figure}[h!]
    \centering
    \resizebox{0.99\textwidth}{!}{
        \includegraphics{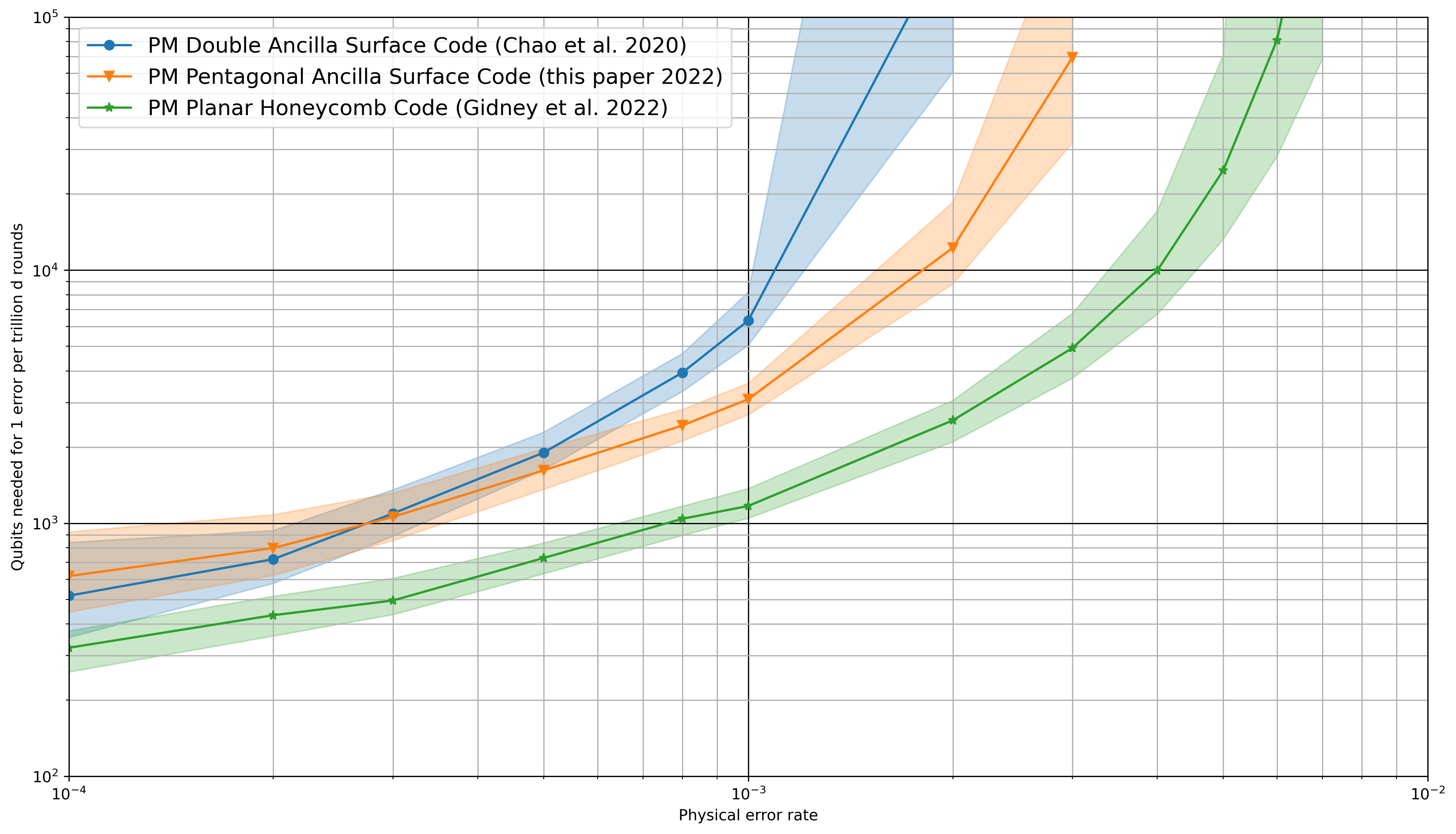}
    }
    \caption{
        Estimated teraquop footprints for error correcting codes compiled into pair measurements.
        Based on X basis memory experiments; see \app{other_plots} for Z basis.
        Derived from the X intercepts at $Y=10^{-12}$ in \fig{extrapolation_plot}.
        The teraquop footprint is the number of physical qubits needed to create a logical qubit large enough to reliably to execute one trillion code distance blocks, which is enough blocks to build classically intractable instances of textbook algorithms like Shor's algorithm.
        See \app{uncertainty} for a discussion of how the line fits used for these data points were computed.
    }
    \label{fig:footprint_plot}
\end{figure}

\fig{footprint_plot} highlights the intercepts of the line fits from \fig{extrapolation_plot} with a target logical error rate of one in a trillion.
In other words, it estimates the number of physical qubits per logical qubit needed to reach algorithmically relevant logical error rates.
In this plot you can see that my construction is initially better than Chao et al's construction, but is improving more slowly as the physical error rate decreases.
This is what I expected, based on the fact that my construction has
bidirectional hook errors in my construction.
Although the data that I collected is not definitive on this point, because of
the bidirectional hook errors I believe that for low enough physical error Chao et al's construction becomes strictly better.
At a physical error rate of 0.1\%, the honeycomb code has a teraquop footprint of around 1000 qubits, compared to 3000 qubits for my construction and 6000 qubits for Chao et al's construction.

\section{Conclusion}
\label{sec:conclusion}

In this paper, I presented a new way to compile the surface code into pair measurements, found by using the ZX calculus.
Although the resulting construction isn't as efficient as the honeycomb code, and although it has bidirectional hook errors that cut its code distance in half, my construction doubled the threshold and halved the teraquop footprint (at a physical error rate of 0.1\%) compared to previous work compiling the surface code into pair measurements.

One of the striking things I noticed, after compiling the surface code into pair measurements, is how the compiled circuit looks a lot like the circuit for a floquet code.
The circuit doesn't really know about the stabilizers that it came from, it only knows about the pair measurements running along the pentagonal faces.
Detectors are formed from the edge measurements by combining them in surprising ways that narrowly avoid anti-commuting with intermediate measurements (see \fig{detector}).
That's also what detectors look like in floquet codes.
Particularly notable is that the compiled circuit is constantly moving the logical observable (for example, see the stim circuits on zenodo~\cite{craig_gidney_2022_6626417} and note the OBSERVABLE\_INCLUDE instructions inside their REPEAT blocks).
Having to constantly move the observable is normally thought of as \emph{the} defining aspect of a floquet code~\cite{hastings2021dynamically}, but here the same property is occurring in a compiled surface code circuit.

Another fact about the circuit that surprised me was how well it performs, despite having half code distance.
In the past, I've tried very hard to make sure I didn't accidentally cut the code distance of a construction in half.
In this paper, I instead just accepted the halved code distance, moved forward, and found that the results were still competitive.
This suggests that (at least for physical error rates above $0.1\%$), whether the code distance was cut in half isn't a reliable predictor of performance.

\section{Acknowledgements}

Thanks to Hartmut Neven for creating an environment where this work was possible.
Thanks to Austin Fowler for writing the correlated minimum weight perfect matching decoder used by this paper.
Thanks to Oscar Higgott for feedback that improved the paper.
Thanks to Matt McEwen for an enormous amount of feedback that improved the paper.

\bibliographystyle{plainnat}
\bibliography{refs}

\appendix
\clearpage

\section{Line Fits}
\label{app:uncertainty}

The line fits shown in \fig{extrapolation_plot} and used for the data points in \fig{footprint_plot} were computed by using a Bayesian method, with a truncation step to represent systemic uncertainty.

The line $Y = mX + b$ was defined to be the hypothesis that, for a qubit count $q$ satisfying $\sqrt{q} = X$, sampling whether or not a shot had a logical error was equivalent to sampling from a Bernoulli distribution with parameter $p=\exp(Y)=\exp(m \sqrt{q} + b)$.

Given a hypothesis, and the data that was actually sampled, you can compute the probability that the hypothesis assigned to seeing that data.
This is the hypothesis' likelihood.
Note that the probability of any one specific thing happening is always tiny, well below the smallest finite values that can be presented by floating point numbers, so likelihood computations have to be done in log space.

The dashed lines shown in \fig{extrapolation_plot} are the max likelihood hypotheses.
They are the lines whose hypotheses assigned the highest probability to the data.
The shaded regions are the union of all the lines whose likelihoods were within a factor of 1000 of the likelihood of the max likelihood hypothesis.

The benefit of this fitting method, over just performing a typical least squares error line fit to the points shown in \fig{extrapolation_plot} are (1) it doesn't require discarding data when no errors are observed, (2) it gives a good sense of the shape of nearby hypotheses, and (3) it naturally accounts for varying statistical uncertainty in the locations of points being used for the fit.

Unfortunately, although this fitting method is essentially optimal at quantifying which hypotheses are more likely, it doesn't deal well with systemic error where the true hypothesis is not one of the hypotheses being considered.
For example, by adding more data, you can make the line fits qualitatively worse.
If you keep collecting more and more samples for two points with high logical error rates, which is cheap to do, you'll force the line fit to exactly predict those two specific points.
But the true system doesn't have logical error rates that lie exactly on a line, so this sacrifices the correspondence to reality of the fit for points at low logical error rates.
And our goal is to predict the behavior at low logical error rates.

To mitigate this problem, I used a hack: I clamped how many errors a point could claim it had.
Any point with $s$ shots and $e$ errors, with $e > 10$, was instead presented to the fitting function as a point with $\lceil s \cdot 10/e \rceil$ shots and $10$ errors.
The truncation is supposed to represent the fact that there is unavoidable uncertainty in how closely the line can estimate the true curve.

Anecdotally, using truncation appears to significantly improve the fits.
It's better than treating all the points equally, which gives low sample count points too much pull (lone "unlucky" errors, which do appear in the data, shift the prediction too drastically).
It's also better than not clamping, which gives high sample count points too much pull and produces obviously overconfident predictions.

Note that \fig{extrapolation_plot} doesn't show data points that had 0 observed errors because, although the points themselves wouldn't be visible, their error bars would all overlap and obscure the data points with less than 5 sampled errors.
Although these data points aren't shown, they were still included when fitting.
The samples they represented correctly affected the likelihoods computed for each hypothesis.

Note that, before line fitting begins, points with sampled logical error rates above 40\% are discarded (to avoid working with points above threshold).

\clearpage
\section{Noise Model}
\label{app:noise_model}

The noise model used by this paper treats all measurements the same: single qubit measurements are treated the same as pair measurements, and demolition measurements are treated the same as non-demolition measurements.
The noise model applies anti-commuting Pauli errors after resets, depolarization after any other operation (including idling), and it probabilistically flips measurement results.
See \tbl{noise_model}.

Note that the model asserts that classically controlled Pauli gates are completely free.
Classically controlled Paulis introduce no noise, they consume no time, they do not prevent other operations from being applied to the target qubit, they don't affect whether the target qubit is idling, and they incur no reaction delay from waiting for measurement results.
This is because, in stabilizer circuits, classically controlled Paulis can always be applied entirely within the control system and can be deferred until later by conjugating them by upcoming Clifford operations.

\begin{table}[h]
    \centering
    \begin{tabular}{|r|l|}
    \hline
    Noise channel & Probability distribution of effects
    \\
    \hline
    $\text{MERR}(p)$ & $\begin{aligned}
        1-p &\rightarrow \text{(report previous measurement correctly)}
        \\
        p &\rightarrow \text{(report previous measurement incorrectly; flip its result)}
    \end{aligned}$
    \\
    \hline
    $\text{XERR}(p)$ & $\begin{aligned}
        1-p &\rightarrow I
        \\
        p &\rightarrow X
    \end{aligned}$
    \\
    \hline
    $\text{ZERR}(p)$ & $\begin{aligned}
        1-p &\rightarrow I
        \\
        p &\rightarrow Z
    \end{aligned}$
    \\
    \hline
    $\text{DEP1}(p)$ & $\begin{aligned}
        1-p &\rightarrow I
        \\
        p/3 &\rightarrow X
        \\
        p/3 &\rightarrow Y
        \\
        p/3 &\rightarrow Z
    \end{aligned}$
    \\
    \hline
    $\text{DEP2}(p)$ & $\begin{aligned}
        1-p &\rightarrow I \otimes I
        &\;\;
        p/15 &\rightarrow I \otimes X
        &\;\;
        p/15 &\rightarrow I \otimes Y
        &\;\;
        p/15 &\rightarrow I \otimes Z
        \\
        p/15 &\rightarrow X \otimes I
        &\;\;
        p/15 &\rightarrow X \otimes X
        &\;\;
        p/15 &\rightarrow X \otimes Y
        &\;\;
        p/15 &\rightarrow X \otimes Z
        \\
        p/15 &\rightarrow Y \otimes I
        &\;\;
        p/15 &\rightarrow Y \otimes X
        &\;\;
        p/15 &\rightarrow Y \otimes Y
        &\;\;
        p/15 &\rightarrow Y \otimes Z
        \\
        p/15 &\rightarrow Z \otimes I
        &\;\;
        p/15 &\rightarrow Z \otimes X
        &\;\;
        p/15 &\rightarrow Z \otimes Y
        &\;\;
        p/15 &\rightarrow Z \otimes Z
    \end{aligned}$
    \\
    \hline
    \end{tabular}
    \caption{
        Definitions of various noise channels.
        Used by \tbl{noise_model}.
    }
    \label{tbl:noise_channels}
\end{table}

\begin{table}[h!]
    \centering
    \begin{tabular}{|r|l|}
    \hline
    Ideal gate & Noisy gate
    \\
    \hline
    (Idling) & $\text{DEP1}(p)$
    \\
    (Single qubit gate) $U_1$ & $U_1 \circ \text{DEP1}(p)$
    \\
    \hline
    $R_X$ & $R_X \circ \text{ZERR}(p)$
    \\
    $R_Y$ & $R_Y \circ \text{XERR}(p)$
    \\
    $R_Z$ & $R_Z \circ \text{XERR}(p)$
    \\
    \hline
    $M_X$ & $M_X \circ \text{MERR}(p) \circ \text{DEP1}(p)$
    \\
    $M_Y$ & $M_Y \circ \text{MERR}(p) \circ \text{DEP1}(p)$
    \\
    $M_Z$ & $M_Z \circ \text{MERR}(p) \circ \text{DEP1}(p)$
    \\
    \hline
    $M_{XX}$ & $M_{XX} \circ \text{MERR}(p) \circ \text{DEP2}(p)$
    \\
    $M_{YY}$ & $M_{YY} \circ \text{MERR}(p) \circ \text{DEP2}(p)$
    \\
    $M_{ZZ}$ & $M_{ZZ} \circ \text{MERR}(p) \circ \text{DEP2}(p)$
    \\
    \hline
    (classically controlled Pauli) $P^m$ & (Done in classical control system.)
    \\
    \hline
    \end{tabular}
    \caption{
        How to turn ideal gates into noisy gates under the "pair measurement depolarizing noise" model used by this paper.
        The model is defined by a single parameter $p$.
        Note $A \circ B = B \cdot A$ meaning $B$ is applied after $A$.
        Noise channels are defined in \tbl{noise_channels}.
    }
    \label{tbl:noise_model}
\end{table}

\clearpage
\section{Additional Data}
\label{app:other_plots}

\begin{figure}[h!]
    \centering
    \resizebox{0.99\textwidth}{!}{
        \includegraphics{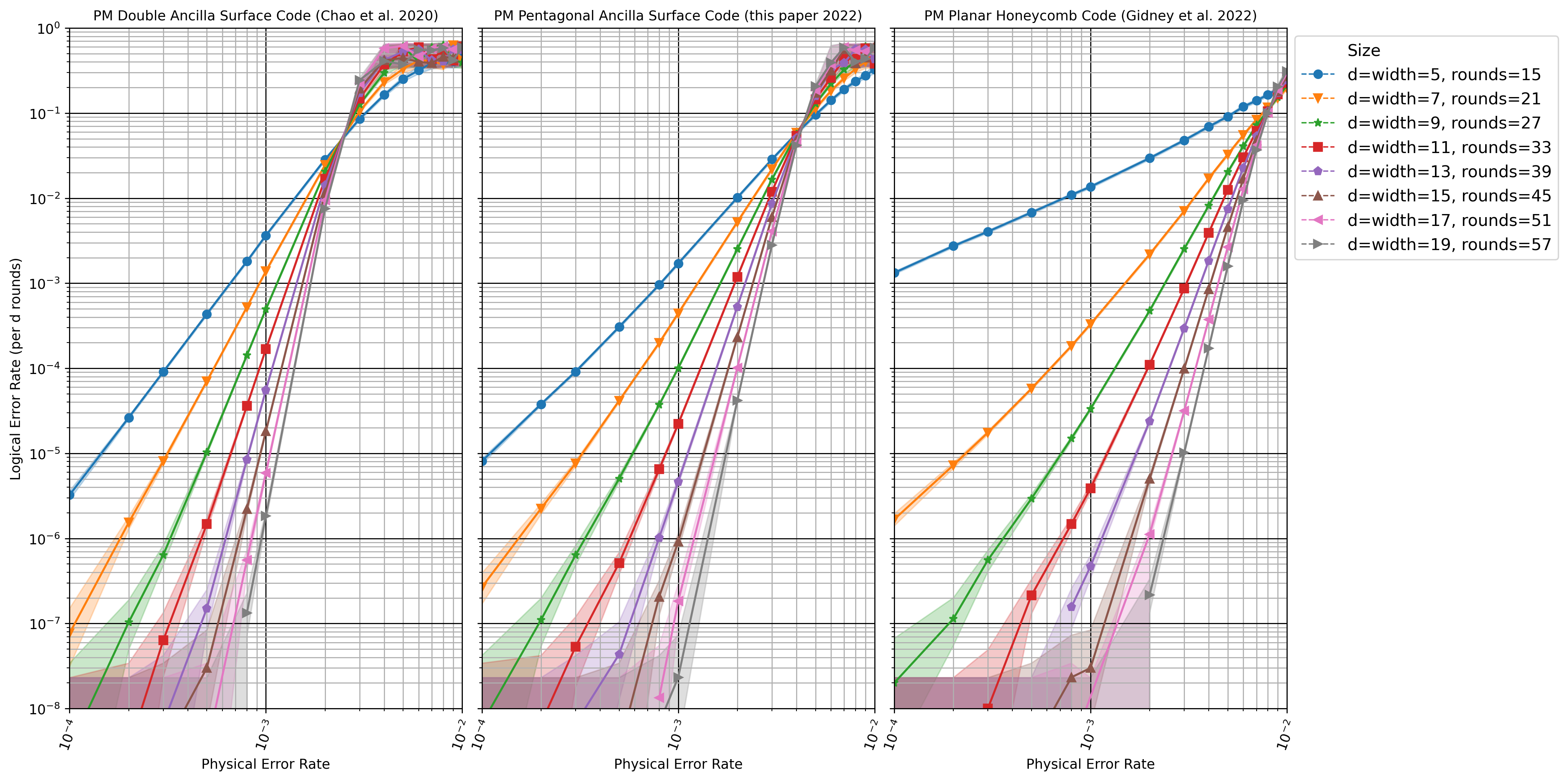}
    }
    \caption{
        Physical error rate vs logical error rate for various patch widths, based on Z basis memory experiments.
        See \fig{error_rate_plot} for X basis.
        The X and Z basis results are not exactly identical (due to differences in the layout of each observable, and due to statistical noise) but are qualitatively identical.
    }
    \label{fig:error_rate_plot_z}
\end{figure}

\begin{figure}[h!]
    \centering
    \resizebox{0.99\textwidth}{!}{
        \includegraphics{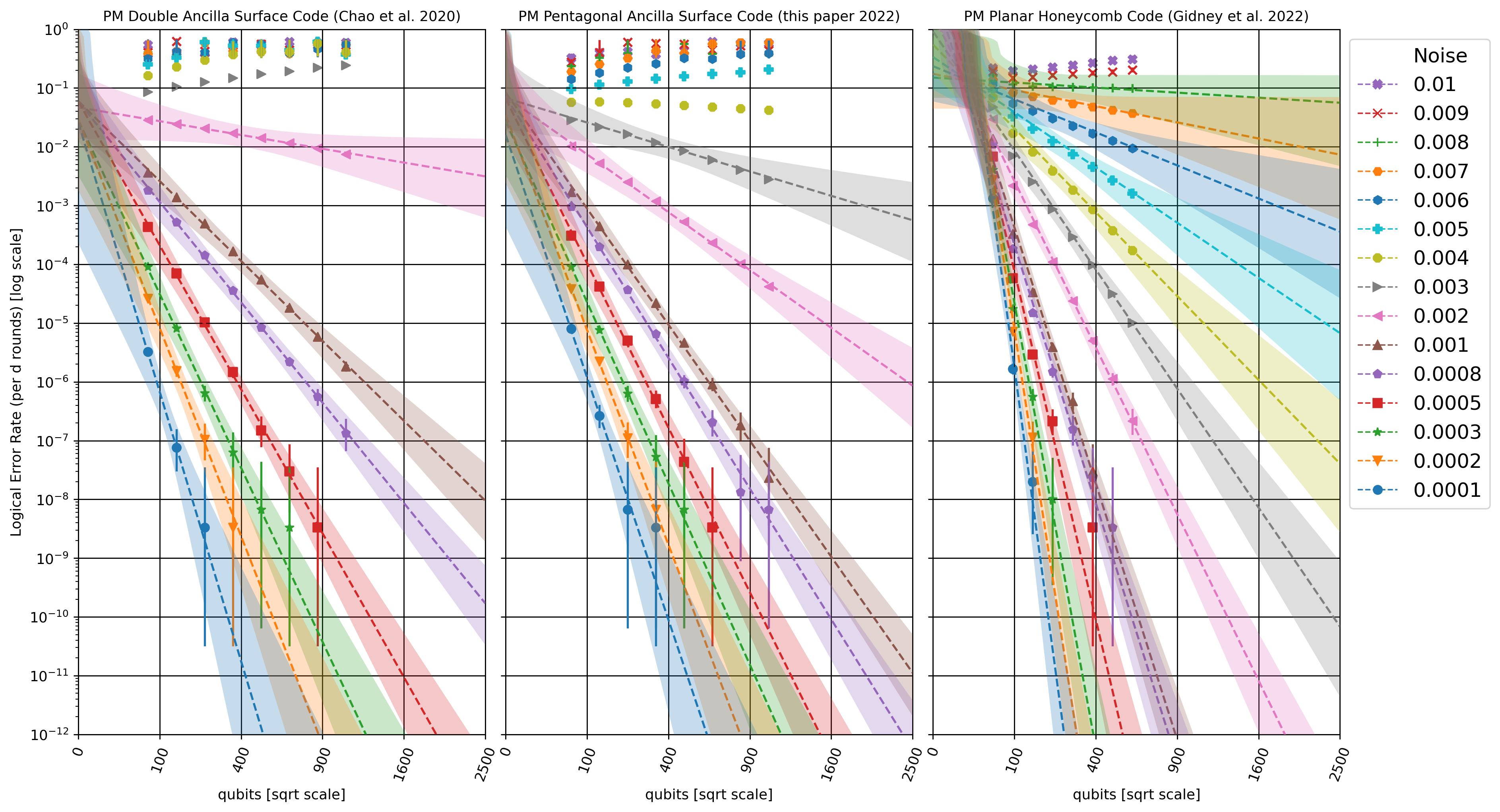}
    }
    \caption{
        Linear extrapolation of log logical error rate versus square root qubit count, for various physical error rates.
        Based on Z basis memory experiments; see \fig{extrapolation_plot} for X basis.
        The X and Z basis results are not exactly identical (due to differences in the layout of each observable, and due to statistical noise) but are qualitatively identical.
    }
    \label{fig:extrapolation_plot_other_z}
\end{figure}

\begin{figure}
    \centering
    \resizebox{0.99\textwidth}{!}{
        \includegraphics{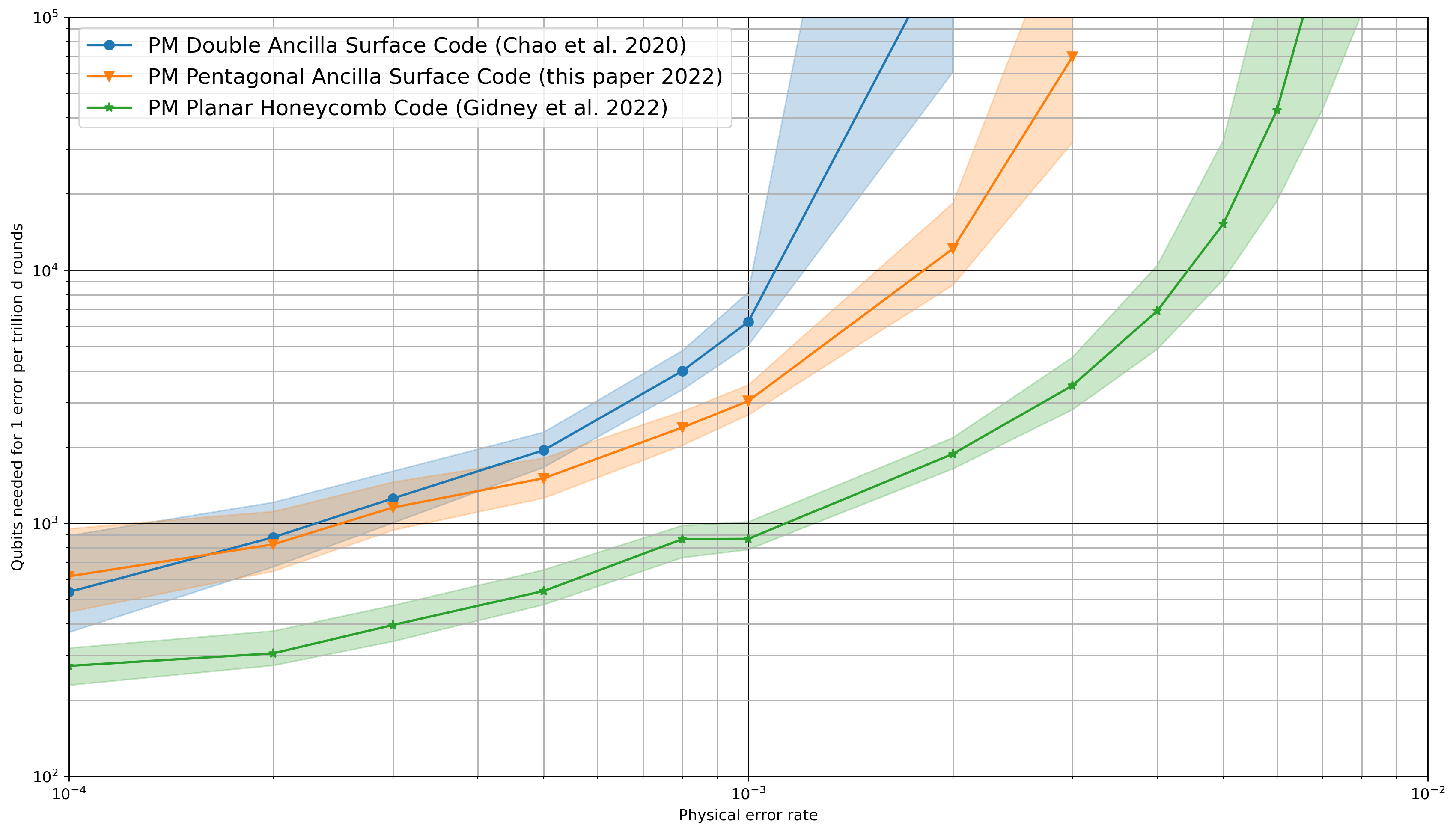}
    }
    \caption{
        Estimated teraquop footprints for error correcting codes compiled into pair measurements.
        Based on Z basis memory experiments; see \fig{footprint_plot} for X basis.
        The X and Z basis results are not exactly identical (due to differences in the layout of each observable, and due to statistical noise) but are qualitatively identical.
    }
    \label{fig:footprint_plot_z}
\end{figure}

\end{document}